\documentclass[]{jfm}

\usepackage{amsmath}
\usepackage{bm}
\usepackage{graphicx}
\usepackage{natbib}
\usepackage{subfigure}
\usepackage{float}
\usepackage{xcolor}
\usepackage{hyperref}
\hypersetup{
    colorlinks = true,
    urlcolor   = blue,
    citecolor  = black,
}
\newcommand{\RomanNumeralCaps}[1]
\newcommand{\br}[1]{\left( #1 \right)}
\newcommand{\dd}{\mathrm{d}}

\title{Departure from the statistical equilibrium of large scales in three-dimensional hydrodynamic turbulence}

\author{Mengjie Ding\aff{1},
 Jin-Han Xie\aff{1,2}
 \corresp{\email{jinhanxie@pku.edu.cn}}
 \and Jianchun Wang\aff{3}}

 \affiliation{\aff{1}Department of Mechanics and Engineering Science at College of Engineering, and State Key Laboratory for Turbulence and Complex Systems, Peking University, Beijing 100871, P. R. China
 \aff{2}Joint Laboratory of Marine Hydrodynamics and Ocean Engineering, Laoshan Laboratory, Shandong 266237, P. R. China
 \aff{3}Department of Mechanics and Aerospace Engineering, Southern University of Science and Technology, Shenzhen 518055, P. R. China}

\begin{document}
\maketitle

\begin{abstract}
We study the statistically steady states of the forced dissipative three-dimensional homogeneous isotropic turbulence at scales larger than the forcing scale in real separation space.
The probability density functions (PDFs) of longitudinal velocity difference at large separations are close to but deviate from Gaussian, measured by their non-zero odd parts.
Under the assumption that forcing controls the large-scale dynamics, we propose a conjugate regime to Kolmogorov's inertial range, independent of the forcing scale, to capture the odd parts of PDFs.
The analytical expressions of the third-order longitudinal structure functions derived from the K\'arm\'an-Howarth-Monin equation prove that the odd-part PDFs of velocity differences at large separations are small but non-zero, and show that the odd-order longitudinal structure functions have a universal power-law decay with exponent $-2$ as the separation tends to infinity regardless of the particular forcing form, implying a significant coupling between large and small scales.
Thus, dynamics of large scales depart from the {\it absolute equilibrium}, and we can partially recover small-scale information without explicitly resolving small-scale dynamics.
The departure from the statistical equilibrium is quantified and found to be viscosity independent.
Even though this departure is small, it is significant and should be considered when studying the large scales of the forced three-dimensional homogeneous isotropic turbulence.
\end{abstract}

\begin{keywords}
	Homogeneous turbulence, Isotropic turbulence
%Authors should not enter keywords on the manuscript, as these must be chosen by the author during the online submission process and will then be added during the typesetting process (see \href{https://www.cambridge.org/core/journals/journal-of-fluid-mechanics/information/list-of-keywords}{Keyword PDF} for the full list).  Other classifications will be added at the same time.
\end{keywords}

%{\bf MSC Codes }  {\it(Optional)} Please enter your MSC Codes here

\section{Introduction}
\label{sec:intro}
In three-dimensional (3-D) homogeneous isotropic turbulence (HIT), the injected energy transfers at a constant rate by nonlinear interactions to small scales until dissipation occurs.
In the inertial range, where the scales are away from the energy-containing scale and the dissipation scale, from the K\'arm\'an-Howarth-Monin (KHM) equation (\citealt{KH38, MY75}), \cite{Kolmogorov41} obtained an asymptotic result for the longitudinal velocity structure function
\begin{equation}
	\langle \delta u_L^3 \rangle = -\frac{4}{5}\epsilon r,
\end{equation}
where $\langle \cdot  \rangle$ denotes the ensemble average, $\delta u_L = (\bm{u}' - \bm{u}) \cdot \bm{r}_0$ is the longitudinal velocity difference with $\bm{u}' = \bm{u}(\bm{x} + \bm{r})$ , $\bm{r} $ the separation vector of two points with $\bm{r}_0=\bm{r}/|\bm{r}|$ a unit vector, and $\epsilon$ is the energy dissipation rate.
The corresponding energy spectrum $E(k) \sim k^{-5/3}$ is also widely observed (\citealt{Obukhov41,Frisch95}), here $k$ is the wavenumber.

For the dynamics of scales larger than the forcing scale (large scales hereafter), since energy cascades from the forcing scale down to small dissipation scales, leaving no averaged upscale energy to the large scales, it has been conjectured that the large-scale dynamics of forced 3-D HIT can be described by the {\it absolute equilibrium} (\citealt{Hopf52,Lee52,RK73,RS78,Frisch95,Lesieur97}).
The spectra of this equilibrium state follow the statistical mechanics of the truncated Euler equations, where energy equally distributes among all Fourier modes and the velocity distribution is Gaussian (\citealt{RS78}).
Based on Liouville's theorem (\citealt{Lee52}) and Gaussian equipartition ensemble (\citealt{Orszag77}),
\cite{RK73} predicted the energy and helicity spectra 
\begin{equation}\label{spec}
	E(k) = \frac{4 \pi \alpha k^2}{\alpha^2 - \beta^2k^2} \quad \mathrm{and} \quad \ H(k) = \frac{4 \pi \beta k^4}{\alpha^2 - \beta^2 k^2},
\end{equation}
respectively.
The coefficients $\alpha$ and $\beta$ are determined by the energy and the helicity of flow.
For non-helical flow, $\beta = 0$, leading to an energy spectrum $E(k) \sim k^2$.
Numerical and experimental results have justified the large-scale statistical equilibrium in 3-D HIT (\citealt{DFA15,CAB17,AB18,AB19,AB20,CBDB05,GF22}).
The dynamics of scales larger than the forcing scale are of interest for many flows where the energy flux is zero, e.g.,
geophysical and astrophysical flow, and turbulent mixing in industrial processes (\citealt{DFA15}).
The large-scale statistical equilibrium is also reported in wave turbulence, such as capillary waves (\citealt{BFLS95,MPF17}), bending waves (\citealt{MNA21}) and optical waves (\citealt{Bau20}).

However, the large-scale dynamics of the forced 3-D HIT differ from the {\it absolute equilibrium} of the truncated Euler equations.
In the {\it absolute equilibrium}, the energy and helicity are conserved in an isolated system, where the initial conditions solely determine the statistical dynamics.
Although the energy flux is zero on average for the large scales of forced 3-D HIT,  \citet{AB19} pointed out that the large-scale dynamics depend on the details of forcing.
Due to the nonlinear interactions, there are transient energy transfers between the large and small scales. Here, the small scales include the forcing, inertial-range and dissipation scales. 
Therefore the spectra quantitatively depart from those predicted by the {\it absolute equilibrium}.

In the free-decaying turbulence, ``Saffman turbulence'' (\citealt{Saffman67}) predicts a similar spectrum, which, however, has different mechanisms from those of the forced turbulence (\citealt{BP56,Davidson15}). In the forced turbulence, averages over statistically steady states are considered, and the large scales interact with small scales. In contrast, the flow is transient in the free-decaying turbulence, and the forcing effect is absent.

An alternative way to distinguish different scales is to consider the separation space (\citealt{DP05}).
This article focuses on the statistics of velocity structure functions at separations larger than the forcing scale.
\citet{GF22} studied the second- and third-order structure functions and found that they are independent of separation $r$ of two measured points, which we believe is a result of low resolution.
This article shows that the probability density functions (PDFs) of velocity difference at large separations are close to but deviate from the Gaussian distribution.
This deviation is measured by the PDFs' nonzero odd part, for which we propose a universal separate-variable form by normalizing the velocity difference using the combination of the downscale energy flux rate $\epsilon$ and the forcing wavenumber $k_f$.
This is a conjugate of Kolmogorov's theory (\citealt{Kolmogorov41}), where the invariant velocity difference PDF in the inertial range is independent of the forcing scale and is obtained after normalizing the velocity by $(\epsilon r)^{1/3}$.
Using the separate-variable-form PDF and the exact expressions for the third-order structure functions, we calculate the analytical expressions for the odd-order structure functions, whose magnitudes' universal slow decay in the limit of large separation implies a strong and robust interaction between large- and small-scale structures.

\section{Theory}
We study the forced 3-D incompressible Navier-Stokes equations
\begin{subequations}
	\begin{align}
		\frac{\partial{\bm{u}}}{\partial{t}} + \bm{u} \cdot \nabla \bm{u} &= - \nabla p + \nu \nabla^2\bm{u} + \bm{F},\\  
		\nabla \cdot \bm{u} &= 0,
	\end{align}
\end{subequations}
where $p$ is the pressure, $\nu$ is the viscosity and $\bm{F}$ is the external forcing.

From the KHM equation (\citealt{Frisch95}), 
%\citet{XB19} derived the theoretical expression of the third-order structure function for the range away from the dissipation scales but including the forcing scales, which is beyond the inertial range. 
we have the relation between the longitudinal third-order structure function and energy injection rate in statistically steady 3-D homogeneous isotropic turbulence:
\begin{equation}\label{S3_e}
	\frac{1}{r^2}\frac{\dd }{\dd r}\br{\frac{1}{r}\frac{\dd }{\dd r}\br{r^4 \langle \delta u_L^3 \rangle}} = - 6{ \langle{\boldsymbol{F}\cdot\boldsymbol{u}'+\boldsymbol{F}'\cdot\boldsymbol{u}} \rangle}.
\end{equation}

Since we focus on the dynamics for scales larger than the forcing scale, let's assume that the forcing effect is confined in the displacement space, i.e., there exists a finite scale $r_c<\infty$ such that the r.h.s. of (\ref{S3_e}) equals $0$ when $r>r_c$, or the forcing effect decays fast enough, e.g., the r.h.s. of (\ref{S3_e}) decays exponentially, in the limit of large displacement, (\ref{S3_e}) gives
\begin{equation}\label{r-2}
	\langle{\delta u_L^3}\rangle \sim r^{-2},
\end{equation}
which is a universal scaling.

Thus, we introduce the type I forcing, which decays exponentially, i.e.,  
\begin{equation}\label{exp_corr}
	\langle \bm{F}\!\cdot\!\bm{F}' \rangle \sim \exp(-(rk_f)^2/4).
\end{equation}
From dimensional analysis, the longitudinal third-order structure can be expressed as
\begin{equation}
		\langle \delta u_L^3 \rangle = \frac{\epsilon}{k_f} g(k_fr).
		\label{eq:S3_general}
	\end{equation}
Substituting (\ref{exp_corr}) into (\ref{S3_e}), the type I forcing leads to  \citep{XB19}
%we obtain the longitudinal third-order structure function \citep{XB19}
%\begin{equation}\label{S3_T1}
%	\langle \delta u_L^3 \rangle = -12\epsilon  \frac{\sqrt{\pi} (z^2-6) \mathrm{erf}(z/2) + 6z\exp(-z^2/4)}{k_fz^4}.
%\end{equation}
%%with
\begin{align}\label{g_I}
	g_{I}(z) = -12\frac{\sqrt{\pi} (z^2-6) \mathrm{erf}(z/2) + 6z\exp(-z^2/4)}{z^4},
\end{align}
%for the type I forcing.
And in the large-scale limit, $r \rightarrow \infty$, we obtain
\begin{equation}\label{g_I_lim}
	\langle \delta u_L^3 \rangle_{I} \rightarrow -12 \sqrt{\pi} \frac{\epsilon}{k_f^3 r^2},
\end{equation}
which is consistent with (\ref{r-2}).

Noting that using Fourier transform we can also express the r.h.s. of (\ref{S3_e}) with the spectral energy injection rate $\epsilon(k)$ as
\begin{equation}
	- 12 \int_{0}^{\infty}\dd k \frac{\varepsilon(k)}{k} \sin(kr),
\end{equation} 
the confined forcing effect in the displacement space also implies a constraint of $\epsilon(k)$. 
Under the constraint of a finite total energy injection rate, an extreme case is associated with a delta concentrated energy injection rate, $\varepsilon(k)=\epsilon\delta(k-k_f)$.
Here, $\varepsilon(k)$ is not an integrable function but is understood as a Dirac measure.
So we introduce the type II forcing with energy injected at a spherical shell, i.e., $|\boldsymbol{k}|=k_f$ with $\boldsymbol{k}$ the wavenumber vector.
Now, solving (\ref{S3_e}), we obtain 
%\begin{equation}\label{S3_T2}
%	\langle \delta u_L^3 \rangle = -12\epsilon  \frac{-z^2\sin(z) - 3z\cos(z)+3\sin(z)}{k_fz^4},
%\end{equation}
\begin{align}\label{g_II}
	g_{II}(z) = -12\frac{-z^2\sin(z) - 3z\cos(z)+3\sin(z)}{z^4},
\end{align}
and it presents a $r^{-2}$ envelope for $\langle{\delta u_L^3}\rangle$ in the limit of large displacement:
\begin{equation}\label{g_II_lim}
	\langle \delta u_L^3 \rangle_{II} \rightarrow 12 \frac{\epsilon}{k_f^3 r^2} \sin(k_f r).
\end{equation}

Combining the above two cases, which we focus on in this paper, we claim that the magnitude of third-order structure function at scales much larger than the forcing scale has a universal $r^{-2}$ scaling.
As an example of other forcing schemes, the results of six-mode forcing are presented in \S \ref{sec_6mode}.

\section{Numerical results}
\label{sec:results}
We employ Direct Numerical Simulations (DNS) using a pseudo-spectral method to test theoretical results.
The Navier-Stokes equations are solved using the pseudo-spectral algorithm in a cubic box with periodic boundary conditions.
The domain size and resolution are $\mathcal{L} = 2\pi$ and $N^3 = 512^3$, respectively.
We use a 2/3 dealiasing and an eighth-order hyper-viscosity for dissipation (\citealt{BO96}).
This algorithm explicitly solves the linear viscous term and uses the Adams-Bashforth method for the nonlinear term (cf. \citealt{CS92}).
We employ type I and II forcings with $k_f = 20$ to an initial weak, random field and collect data in statistically steady states.

\begin{figure}
	\includegraphics[width =\linewidth]{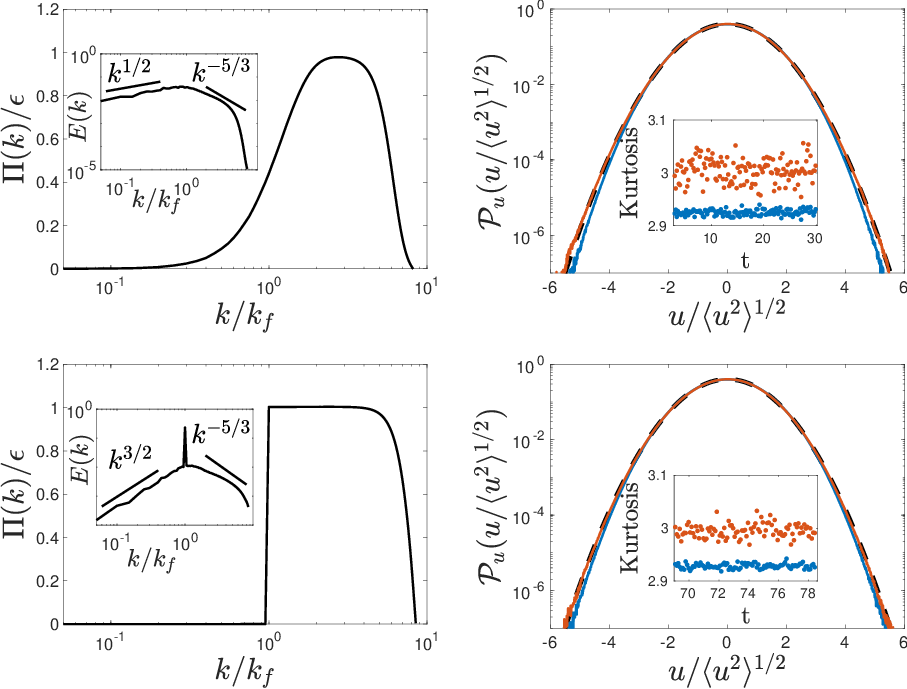}
	\caption{\label{fig:spectra} Left: The normalized energy flux $\Pi(k)/\epsilon$ with inset the energy spectrum $E(k)$.
		Right: PDFs of the normalized fluid velocity fluctuations $\mathcal{P}(u/\langle u^2 \rangle^{1/2})$ of all modes (blue) and large-scale modes (orange). The dashed black line denotes the Gaussian distribution with inset the corresponding Kurtosis. 
		The upper and lower rows represent the exponential (type I) and spherical shell (type II) forcing, respectively.}
\end{figure}

The left panels of Fig. \ref{fig:spectra} shows the energy spectrum $E(k)$ and the energy flux across scales $\Pi(k)$.
For scales smaller than the forcing scale ($k/k_f > 1$), an inertial range associated with a forward energy cascade and a $k^{-5/3}$ energy spectrum is observed, which are consistent with Kolmogorov's theory.
The averaged energy flux is zero for the scales larger than the forcing scale ($k/k_f < 1$), referred to as the no-flux range.
The energy spectra scale as $k^{3/2}$ and $k^{1/2}$ for two forcing types, respectively.
These apparent deviations from the $k^2$ spectrum of the {\it absolute equilibrium} state are related to the nonlinear interactions between large scales and forcing scale, which are also observed and explained in \cite{AB19}.

We obtain the velocity fluctuations of large-scale modes by applying a spectral low-pass filter, which sets the modes with $k>k_f$ to zero.
In the right panels of Fig. \ref{fig:spectra}, PDFs of the normalized velocity fluctuations $ \mathcal{P}_u(u/\langle u^2 \rangle^{1/2})$ of large-scale modes are shown to be close to Gaussian with Kurtosis near $3$.

\begin{figure}
	\includegraphics[width =\linewidth]{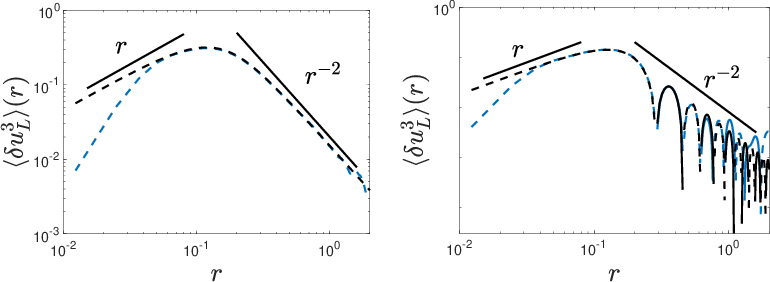}
	\caption{\label{fig:deltaul}Theoretical solution (black) and results from DNS (blue) for the longitudinal third-order structure function $\langle \delta u_L^3 \rangle$. The left and right panels represent the exponential (type I) forcing and spherical shell (type II) forcing, respectively.
		Solid lines represent the positive values, and dashed lines represent the absolute value of the negative values.}
\end{figure}

We check in Fig. \ref{fig:deltaul} that the theoretical expressions (\ref{eq:S3_general}), (\ref{g_I}) and (\ref{g_II}) well capture the longitudinal third-order structure function $\langle \delta u_L^3 \rangle$ obtained from DNS in the inertial range, forcing scale and the no-flux range (large scales).
In the inertial range, $\langle \delta u_L^3 \rangle$ scales as $r$, consistent with Kolmogorov's theory.
In the no-flux range, $r > l_f$ with $l_f=\pi/k_f$ the forcing scale, as $r \rightarrow \infty$, the envelope of $\langle \delta u_L^3 \rangle$ scales as $r^{-2}$.
Since we employed a triple periodic boundary condition, the theoretical and numerical results collapse to a range smaller than the domain size.

While the dynamics of large scales depend on the details of forcing and large-scale spectrum alters, 
the longitudinal third-order structure function $\langle \delta u_L^3 \rangle$ has a nonzero behaviour independent of the form of forcing.
This implies that the interactions between large scales and small scales are robust.
%indicating that the large and small (forcing and energy cascades) scales are well coupled. 
We also note that in the first case while the forcing decays exponentially with separation $r$, the third-order structure function still decays slowly as $r^{-2}$, suggesting that the interactions are vigorous.
In \S \ref{sec_6mode} we justify the decay rate with data from numerical simulations with six-mode forcing, which is suggested by \cite{AB19} to generate $k^2$ spectrum at large scales.

\begin{figure}
	\includegraphics[width =\linewidth]{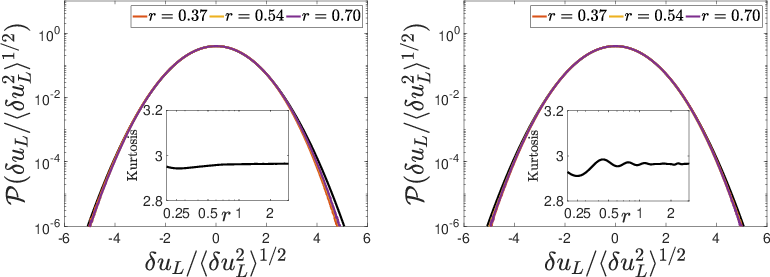}
	\caption{\label{fig:pdf_deltau}
		PDFs of the normalized longitudinal velocity difference $\mathcal{P}(\delta u_L / \langle \delta u_L^2 \rangle ^{1/2})$ at large scales (greater than the forcing scale $r_f \approx 0.16$). The black line represents the Gaussian distribution. Inset is the Kurtosis of $\delta u_L / \langle \delta u_L^2 \rangle ^{1/2}$ at large separations $r$. The left and right panels represent the exponential (type I) and spherical shell (type II) forcing, respectively.}
\end{figure}

In Fig. \ref{fig:pdf_deltau}, we present the PDF of the normalized longitudinal velocity difference  $\mathcal{P}(\delta u_L / \langle \delta u_L^2 \rangle ^{1/2})$ with separations $r = 0.37$, $0.54$, and $0.70$, which correspond to the local extrema of the third-order structure function expression divided by the decaying rate $r^{-2}$ (cf. (\ref{g_II_lim})) and are larger than the forcing scale $r_f = \pi/k_f \approx 0.16$. 
The observed PDFs are close to Gaussian distributions, and we quantify the small departure from Gaussian using the odd parts of the PDFs, which are shown in Fig. \ref{fig:pdf_deltau2} and are discussed with details in the following subsection.
%Here, we focus on the odd-order structure functions and the odd part of the PDF, whose nonzero behaviours are clear measures of the departure from Gaussian.

%The 2/3 rule is used to remove aliasing errors.
%Eighth-order hyper-viscosity is used to dissipate energy transferred to small scales \cite{BO96}.

%We employed a modified Euler method to start(or restart) the simulations.
%The code used is described in detail in literature .
%The forcing is a solenoidal, Gaussian random field and delta-correlated in time.

%We maintain the turbulent flow with constant injected energy for several large-eddy turnover times and 
%(see Sec. I of Supplementary Materials) \com{Sec I in SM does not explain this deviation, but shows results with another type of forcing}.
%	\vspace{2ex}
%	\hspace{2ex}
%	\includegraphics[width = 0.35 \textwidth]{pdf_u.eps}
%	\includegraphics[width = 0.35 \textwidth]{Spec_Flux_20fe.eps}
%	\vspace{2ex}
%	\hspace{2ex}
%	\includegraphics[width = 0.35 \textwidth]{pdf_u_20fe.eps}
%We also present in Sec. I of Supplementary Materials that when energy is injected in six Fourier modes ($\pm k_f$,0,0), (0,$\pm k_f$,0), (0,0,$\pm k_f$), $\langle \delta u_L^3 \rangle$ also decays as $r^{-2}$.
%Sec. II of Supplementary Materials also shows that $\langle \delta u_L^3 \rangle\sim r^{-2}$ with six-mode energy injection.

\subsection{Expressions of PDF and high-odd-order structure functions}
Since the large and forcing scales are well coupled, we hypothesize that the forcing mechanism dominates the dynamics of the no-flux regime (large scales).
Therefore the odd part of the probability distribution of velocity difference at large separations has a separate-variable form where the velocity difference is normalized by the combination of characteristic wavenumber $k_f$ and energy injection rate $\epsilon$:
%\com{do we have to say this?},
\begin{equation}\label{p_odd}
	\mathcal{P}_{odd}\left(\delta u_L, \epsilon, k_f, r\right) = \mathcal{P}_0\left(\delta u_L/\left({\epsilon}/{k_f}\right)^{1/3}\right) g(k_f r).
\end{equation}
This expression is quite remarkable as a conjugate of Kolmogorov's theory, where the inertial-range PDF of velocity difference is normalized following $\delta u_L/(\epsilon r)^{1/3}$.

Thus, in the no-flux range, the odd-order structure functions are obtained from a direct integration using (\ref{p_odd})
\begin{equation}
	\begin{split}
		\langle \delta u_L^n \rangle = \int \delta u_L^n \mathcal{P}_{odd}(\delta u_L, \epsilon, k_f, r) d\delta u_L  =C_n \left(\frac{\epsilon}{k_f}\right)^{\frac{n}{3}} g(k_f r),
	\end{split}
	\label{nf-sf2}
\end{equation}
where
\begin{equation}
	C_n = \int z^n \mathcal{P}_0(z) dz,
\end{equation}
$n$ is an odd integer, and $g(k_fr)$ is universal and can be obtained from the already calculated analytical third-order structure function expressions (cf. (\ref{g_I}) and (\ref{g_II})).

\begin{figure}
	\includegraphics[width =\linewidth]{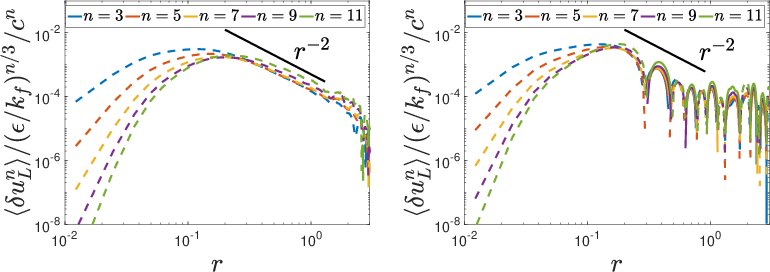}
	\caption{\label{fig:deltaul2} Normalized high-odd-order structure function $\langle \delta u_L^n \rangle / (\epsilon / k_f)^{n/3}/c^{n}$.
		The left and right panels represent the exponential (type I) forcing ($c=7.0$) and spherical shell (type II) forcing ($c=6.6$), respectively.
		Solid lines represent the positive values, and dashed lines represent the absolute value of the negative values.}
\end{figure}

\begin{figure*}
	\centering
	\includegraphics[width =0.9\textwidth]{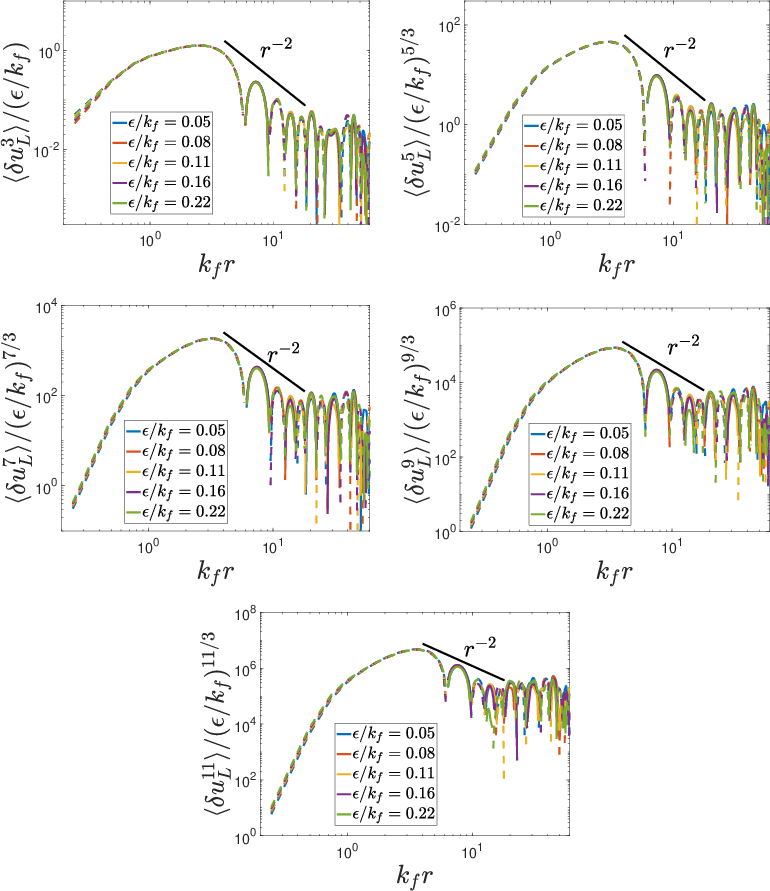}
	\caption{Normalized high-odd-order structure functions in numerical simulations with type II forcing and varying energy injection rates.}
	\label{fig:high_epskf} 
\end{figure*}

In Fig. \ref{fig:deltaul2}, we justify this expression up to order 11 and find that $C_{n} \approx -12 c^{n-3}$, with $c = 7$ and $c=6.6$ for the two forcing types, respectively.
The convergence of the high-order structure functions is shown in \S \ref{appA}.
Even though the type I forcing decays rapidly with separation $r$, the odd-order structure functions decay slowly ($r^{-2}$), resulting from the vigorous interactions between large and small scales. 
%and {\color{red}$C_{n} \approx -12\ 6.6^{n-3}$ which is obtained numerically}. 
This expression is further justified by the collapse of structure functions with different dissipation rates $\epsilon$ using the spherical shell (type II) forcing in Fig. \ref{fig:high_epskf}.

\begin{figure}
	\includegraphics[width =\linewidth]{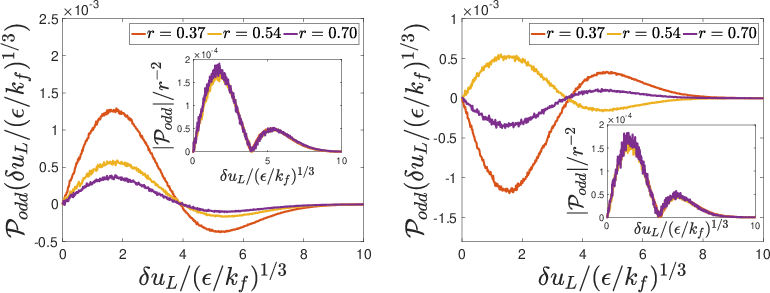}
	\caption{\label{fig:pdf_deltau2}
		Odd part of the PDF of normalized $\delta u_L$ by $(\epsilon/k_f)^{1/3}$ at large separations and the absolute value of that divided by $r^{-2}$ (inset). The left and right panels represent the exponential (type I) and spherical shell (type II) forcing, respectively.}
\end{figure}

We justify the separate-of-variable form of the PDF (\ref{p_odd}) in the insets of Fig. \ref{fig:pdf_deltau2}.
When the PDFs are normalized by $g(k_f r)$, the curves collapse well, suggesting a universal $\mathcal{P}_0$.
For the no-flux range (large scales), $g(k_f r) \sim r^{-2}$ for turbulence driven by type I and II forcing.
The $r$-dependence of $\mathcal{P}_{odd}$ represent the forcing effect, along with the universal of $\mathcal{P}_0$ indicates that the forcing dominates the dynamics of large scales.
Thus, even though the non-Gaussian odd-part PDF of the longitudinal velocity difference, $\mathcal{P}_{odd}$, is small ($O(10^{-3}$)), it implies a strong and robust interaction between large and small scales.

%For example, 
%\begin{equation}
%g(k_fr) = \frac{-k_f^2r^2\sin(k_fr) - 3k_fr\cos(k_fr)+3\sin(k_fr)}{k_f^4 r^4},
%\end{equation}
%and
%\begin{equation}
%g(k_fr) = \frac{\sqrt{\pi} (k_f^2r^2-6) erf(k_fr/2) + 6k_fr\exp(-k_f^2r^2/4)}{k_f^4 r^4},
%\end{equation}
%corresponding to the two forcing types (cf. (\ref{eq:S}) and (\ref{eq:S_20fe})), respectively.
%Normalized high-order structure functions $\langle \delta u_L^n \rangle / (\epsilon / k_f)^{n/3}$ with $n = 3, 5, 7, 9, 11$.
%The theoretical expression of the high-order structure functions at large scales when the flow is forced on the modes inside a finite-width spherical shell is proposed as {\color{red}(cf.  \ref{nf-sf2} in the main text.)}
%\begin{equation}
%\begin{aligned}
%	&\langle \delta u_L^{n} \rangle =
%	&C_{n}\left(\frac{\epsilon}{k_f}\right)^{n/3} \frac{-k_f^2r^2\sin(k_fr) - 3k_fr\cos(k_fr)+3\sin(k_fr)}{k_f^4 r^4},
%\end{aligned}
%\end{equation}

\section{Summary}
In summary, we study the large-scale statistics of the forced 3-D homogeneous isotropic turbulence in the real separation space and find that the PDFs of the longitudinal velocity difference are close to but depart from a Gaussian distribution. We measure this departure by the odd-part PDFs.
From the argument that large scales and forcing scale are well coupled, the large-scale dynamics ($r > r_f$) are dominated by the forcing mechanism, the odd-part PDF of velocity difference $\mathcal{P}_{odd}$ for large separations can be written to a separate-variable form by separating the $r$-dependence, which leads to a universal $r$-dependence for odd-order structure functions. 
Even though the odd part of the PDF has a smaller magnitude than the even part, the analytical solutions of the third-order structure function calculated from the KHM equation prove that the odd-part PDF is nonzero and the $r$-dependence of odd-order structure functions decays slowly following a power-law $r^{-2}$ as $r \rightarrow \infty$, implying that the long-range interaction between large and small scales is strong and robust.
%\textcolor{red}{Thus, the departure is quantified and viscosity independent. It is significant despite the fact that it is small and should be considered studying the large scales of forced 3-D HIT .}
Interestingly, this large-scale regime with a characteristic scale given by the forcing scale is conjugate to the inertial range of Kolmogorov's theory, where the characteristic scale is the separation.
This discovery provides a more comprehensive understanding of turbulence because previous turbulence theory focuses on the inertial range, which cannot exist solely. And we need an understanding of dynamics at scales larger than the forcing scale.

An interesting implication is that if we have an observation of 3-D turbulence which cannot resolve the forcing scale,
we can still know if there is unresolved energy injection, which could be a foundation for turbulence superresolution.
This implication is reasonable: even though there is no averaged upscale energy flux, the large-scale field needs to adjust accordingly to reach a statistically steady state with a forward cascade.
From another perspective, if the 3-D turbulence is driven from an initially zero state, a transient energy flux to large scales must exist before reaching the statistically steady state, which links the small and large scales through energy injection and is remembered by the statistically steady state.

\begin{acknowledgments}
M. Ding and J.-H. X. acknowledge financial support from the National Natural Science
Foundation of China (NSFC) under grant no. 92052102 and 12272006, and by the Laoshan Laboratory under grant no. 2022QNLM010201.
J. Wang acknowledges financial support from the NSFC under grant no. 12172161.
\end{acknowledgments}

\appendix

\section{Six-mode forcing}\label{sec_6mode}

\begin{figure}
	\centering
	\includegraphics[width =0.9\textwidth]{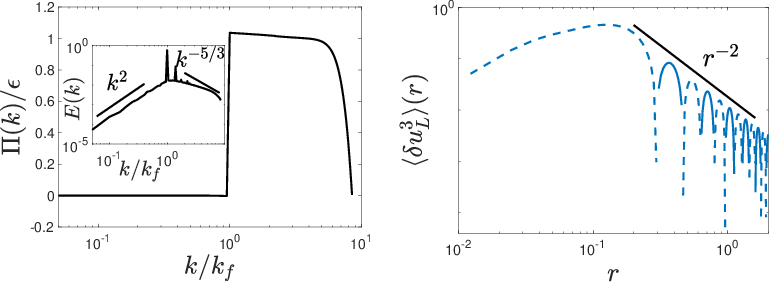}
	\caption{\label{fig:20fs} Results for forcing on six modes: ($\pm k_f$,0,0), (0,$\pm k_f$,0), (0,0,$\pm k_f$). Left: spectra and energy flux. Right: the third-order longitudinal structure function. Solid lines represent the positive values, and dashed lines represent the absolute value of the negative values.}
\end{figure}

As suggested in \cite{AB19}, we performed a simulation where forcing is added on six modes: ($\pm k_f$,0,0), (0,$\pm k_f$,0), (0,0,$\pm k_f$). 
In this case, the forcing effect on the dynamics of large scales is weaker compared with type I forcing (exponential forcing) and type II forcing (spherical shell forcing).
In the no-flux range of Fig. \ref{fig:20fs}, the spectrum has $k^2$ scaling, consistent with the {\it absolute equilibrium}.
The longitudinal structure function, however, decays slowly as $r^{-2}$, the same as the other two forcing methods.

\section{\label{appA}  Convergence of the PDF of the normalized longitudinal velocity difference}
\begin{figure*}
	\centering
	\includegraphics[width =0.7\textwidth]{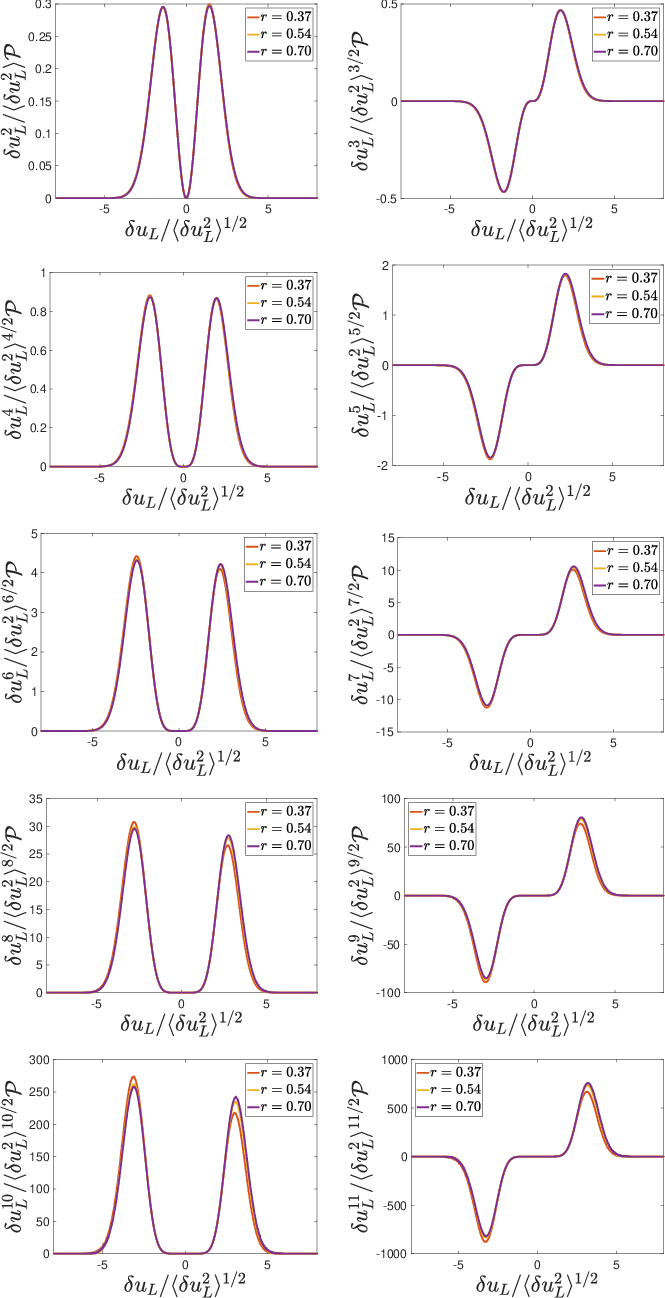}
	\caption{\label{fig:pdf_con3}$\mathcal{P}(\delta u_L / \langle \delta u_L^2 \rangle^{1/2})$ multiplied by $\delta u_L^n / \langle \delta u_L^2 \rangle^{n/2}$ ($n=2,3,4,5,6,7,8,9,10,11$) at different scales for the exponential (type I) forcing.}
\end{figure*}

\begin{figure*}
	\centering
	\includegraphics[width =0.7\textwidth]{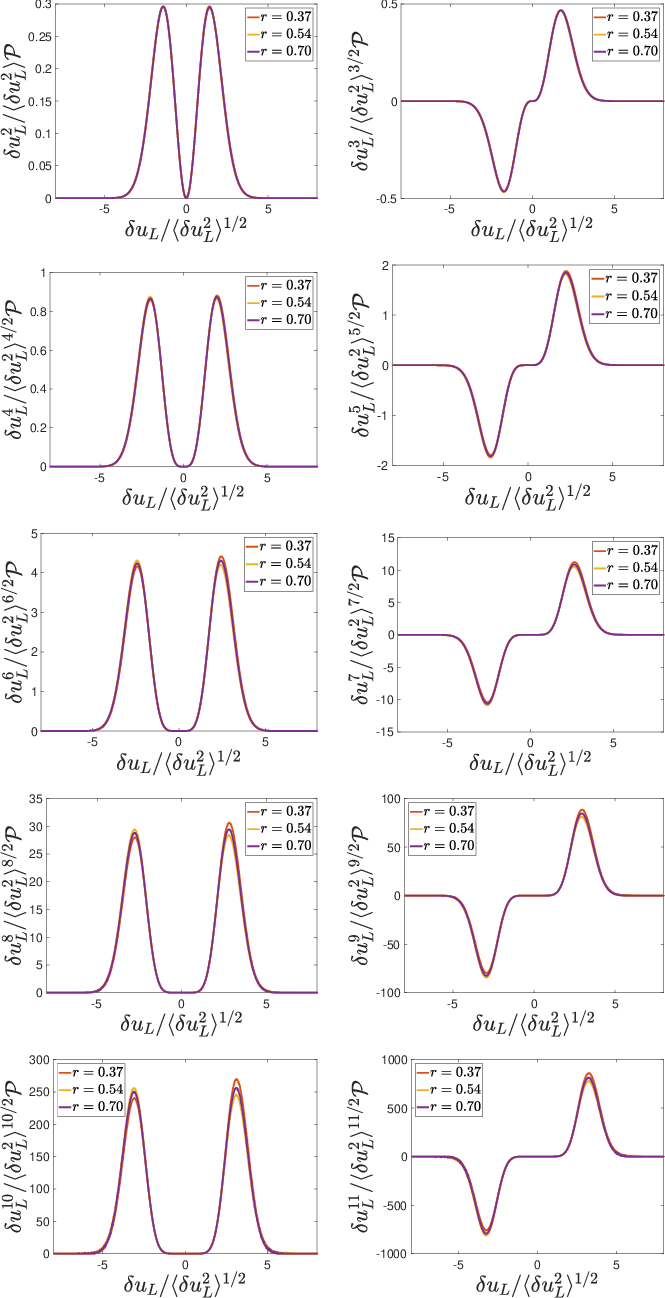}
	\caption{$\mathcal{P}(\delta u_L / \langle \delta u_L^2 \rangle^{1/2})$ multiplied by $\delta u_L^n / \langle \delta u_L^2 \rangle^{n/2}$ ($n=2,3,4,5,6,7,8,9,10,11$) at different scales for the spherical shell (type II) forcing.}
        \label{fig:pdf_con1}
\end{figure*}

The PDFs of the normalized longitudinal velocity difference $\mathcal{P}(\delta u_L / \langle \delta u_L^2 \rangle^{1/2})$ are well resolved as shown in Fig. \ref{fig:pdf_con3} for the exponential (type I) forcing and Fig. \ref{fig:pdf_con1} for the spherical shell (type II) forcing.

\bibliographystyle{jfm}
\bibliography{references}

%% End of file `jfm2esam.bib'.

\end{document}